\documentstyle[twocolumn,aps,prl,epsf]{revtex}
\newcommand{\avg}[1]{\langle{#1}\rangle}

\begin{document}
\draft
\wideabs
{
%\twocolumn[\hsize\textwidth\columnwidth\hsize\csname@twocolumnfalse\endcsname
\title{Measures of globalization based on cross-correlations of world financial indices.}
\author{Sergei Maslov}

\address{Department of Physics, Brookhaven National Laboratory,
Upton, New York 11973, USA}
\date{\today}
\maketitle

\begin{abstract}
The cross-correlation matrix of daily returns of stock market indices in
a diverse set of $37$ countries worldwide was analyzed. Comparison of
the spectrum of this matrix with predictions of random matrix theory provides
an empirical evidence of strong interactions between individual
economies, as manifested by three largest eigenvalues and the corresponding
set of stable, non-random eigenvectors. The observed correlation
structure is robust with respect to changes in the time horizon
of returns ranging from 1 to 10 trading days, and to replacing
individual returns with just their signs. This last observation
confirms that it is mostly correlations in signs and not absolute values of fluctuations,
which are responsible for the observed effect. Correlations between different trading
days seem to persist for up to 3 days before decaying to the level of the
background noise.
\end{abstract}
%\twocolumn]
}
\narrowtext

%\section{Introduction}

In spite of the tremendous importance that current public opinion places
on issues of globalization of the world's economy, its sources and consequences
remain poorly understood. Large downturns and collapses of the economic
and financial situation in one country are routinely blamed on recent
events in other countries. This point of view is reinforced by sensational newspaper
headlines like ``Latin American markets catch the asian flu''.
The level of globalization of a diverse set of 50
developed countries and key emerging markets worldwide
was recently measured and reported in the A.T. Kearney/Foreign Policy
Magazine Globalization Index$^{\rm TM}$ \cite{kearney}.
The factors selected to contribute to this index are extremely diverse and
include among other things volumes of inward- and outward-directed
foreign investments, the amount of international travel and phone calls,
number of servers of the World Wide Web, etc.
Among other things globalization is expected to manifest itself in the dynamics of
financial indices of stock markets in different countries. Indeed, it is reasonable
to expect that a significant coupling of the economy of a given country to the
rest of the world (e.g through foreign investments), would make
its stock index more susceptible to changes in the world economic climate.
%This index along with
%the Foreign Direct Investment (FDI) Confidence Index$^{\rm TM}$ were
%produced by a well known global management-consulting company and are widely used
%in making decisions about foreign investments.

In this work we suggest a simple measure of the level of financial
globalization of a given country based on the analysis of cross-correlations
between stock market indices in different countries and regions of the world.
The main object of our study is the $N \times N$ empirical
correlation matrix $C_{ij}$ of index price fluctuations
in a large number of individual countries ($N=37$ in our study).
The matrix is constructed by applying the formula
$C_{ij}={1 \over T}\sum_{t=1,T} \delta x_i (t) \delta x_j (t)$ to the set
of normalized local currency returns of individual indices, recorded over
a period of $T$ trading days. A return $\delta X_i(t)$ of the stock
index $S_i(t)$ with the time horizon $\Delta t$ is usually defined as
$\delta X_i(t)=\ln S_i(t+\Delta t)-\ln S_i(t) \simeq ( S_i(t+\Delta t)-S_i(t))/ S_i(t)$.
Different markets are characterized by different volatilities of their stock market
indices. In order to be able to detect similarity in the pattern of returns
in different countries on needs to exclude volatility effects by using
normalized returns $\Delta x_i(t)$. $\Delta x_i(t)$ is constructed by
offsetting each $\delta X_i(t)$ by its empirical average value $\avg{\delta X_i}$,
and normalizing it by its empirical variance (volatility):
$\delta x_i(t)={\delta X_i(t)-\avg{\delta X_i} \over \sqrt{\avg{\delta X_i^2}-\avg{\delta X_i}^2}}$.
The matrix $C_{ij}$ defined using $\delta x_i(t)$ has the property that
in the absence of correlations and in the limit $T \gg N$ it is just the unity matrix.
However in real life the number of trading days $T$ in one's dataset is always
finite. As a result an empirically measured matrix $C_{ij}$ is always dressed by a
substantial amount of noise. It is exactly the task of separating any real correlations
present in the signal from this spurious noise, that makes the analysis of real world
data highly non-trivial. Mainstream economics literature was mostly devoted to a detailed analysis
of these correlations for just a pair of stock indices, e.g. those of New York and
Tokyo Stock Exchanges \cite{econo-us-japan}, but even if a large correlation matrix was considered
usually little effort was made to reliably separate the signal from the noise.
As is common in statistics, the job of looking for correlation patterns
among several noisy signals gets much simpler when the number of signals is large.
In our data this corresponds to a large number of country indices $N$.
The spectral analysis of the correlation matrix followed by a comparison
of the spectrum with predictions of the Random Matrix Theory is a useful tool
which allows one to detect even weak correlations between multiple signals.
This method was recently successfully applied towards finding
reproducible correlations between price fluctuations
of hundreds of stocks traded on the stock exchange of a single
country \cite{bouchaud}, and two countries \cite{drozdz}.
In this work we go one step further and apply the techniques pioneered in
\cite{bouchaud} to a large number of
stock market indices in a geographically diverse set of countries.
An alternative method of analysis of the financial data
from the matrix of correlation coefficients by constructing the
Minimal Spanning Tree (MST) was described in \cite{mantegna},
and recently applied to cross-correlations of world financial indices
in \cite{bonanno}. Authors of this work also found a non-trivial
correlation pattern manifested by a strong regional grouping of
indices in the MST.

The raw data we had at our disposal consists of the daily open, high, low, and close
prices of leading market indices (one per country) in 15 European, 14 Asian, and 8
North and South American countries. We first calculated the daily open-to-close
returns of each of these indices. We further selected from our set only those trading
days for which we had a valid record for each and every country on our
list. All data have gaps in them e.g. due to national holidays, when
a particular market was closed. That left us with precisely 226 trading days
approximately uniformly distributed between April 28, 1998 and December 20, 2000.
Each of these remaining daily returns was normalized in such a way that
$\sum_{t=1}^{226} \delta x_i(t)=0$, and $\sum_{t=1}^{226} \delta x_i(t)^2=1$.
The histogram of all 37 eigenvalues of the correlation matrix $C_{ij}$,
shown in Fig. 1(a), revealed that the majority of
eigenvalues are consistent with a null hypothesis of independent identically distributed
Gaussian variables $ \delta x_i (t)$. The prediction for the eigenvalue density
\begin{equation}
\rho_{RMT}(\lambda)={T \over 2 \pi \sigma^2 N} {\sqrt{(\lambda-\lambda_{-})
(\lambda_{+}-\lambda)} \over \lambda} \,
\end{equation}
given for this situation by Random Matrix Theory \cite{RMT}
reasonably agrees with our data below $\lambda \simeq 1.3$.
This formula, derived in the limit of very large $T$ and $N$, predicts
sharp lower and upper cutoffs, $\lambda_{\pm}=\sigma^2(1+(N/T)\pm 2\sqrt{N/T})$, in the
eigenvalue density. This gives a strict quantitative test for deciding whether
a particular eigenvalue reflects a real correlation signal present in the
data, or is just a spurious noise effect caused by the finite length $T$
of the data set. In principle, any eigenvalue significantly above
the upper cutoff $\lambda_{+}$ should be treated as signal.
The variance $\sigma$ of $\delta x_i(t)$ can be renormalized
from its starting value $\sigma=1$ by the presence of correlations in the data.
Indeed, we obtain the best fit of the noise-band part of the spectrum for
${\tilde \sigma}^2=0.67$, consistent with the empirically observed correlations.
The three largest eigenvalues $2.2, 3.5$, and $8.7$ sufficiently
exceed the theoretical upper limit $\lambda_{+}=1.97$ to be attributed to true
correlation patterns. Indeed, in a control test we found that the probability
that the largest eigenvalue generated by an uncorrelated univariate Gaussian signal
of the same dimensions as our data to exceed 2.2 is around 0.05\%. This should be
contrasted with a typical 5\% to 1\% confidence level of correlations
between a pair of individual indices reported e.g. in \cite{econo-us-japan}.
In order to check the reproducibility of largest eigenvalues and their
corresponding eigenvectors we divided ours set into two consecutive
$113$-point subsets and repeated our analysis. The
existence of 3 outlier eigenvalues did not change, however their values
have slightly changed. The largest eigenvalue was measured to be $9.9$
during the first time interval, and $7.8$ during the second.
As can be concluded from  the inset to Fig. 1,
and Fig. 2, the corresponding eigenvectors are remarkably stable with
overlaps between eigenvectors for the first and the second subintervals
being $0.95$, $0.81$, and $0.67$ for the largest,
the second, and the third eigenvalues correspondingly. The largest
possible overlap, realized when two eigenvectors are identical, is equal to 1.
On the other hand, overlaps between eigenvectors from the noise-band between
$\lambda_{-}$ and $\lambda_{+}$ seem to be purely random (see inset to Fig. 1).
Similar effects but with larger number of outliers above $\lambda_{+}$ (up to around 25) were observed
\cite{bouchaud} for individual stocks traded on US stock exchanges.

Components of the three highest ranking eigenvectors, measured for
our data both in its entirety, and when divided in two equal subintervals, are given in
Table 1 and plotted in Fig. 2. The first interesting result is that
virtually all components of the largest eigenvector are
positive, which means that there are no indices which are anti-correlated with
others.
%\footnote{ We observed it to be true also for daily and weekly
%close-to-close returns of indices. However, when in attempt to investigate the
%effects of time-zone differences, we shifted all Asian indices
%one day ahead of European and American ones, the components of the highest
%ranking eigenvector for Asian markets became negative, indicating the appearance
%of anti-correlations}.
 Since eigenvectors corresponding to different eigenvalues
have to be orthogonal to each other, other eigenvectors must contain negative
components. The first (largest) eigenvector has strong support
in European and American sectors, while its components in the
Asian sector are somewhat smaller (yet still positive).
The second eigenvector, on the other hand, is largely dominated by Asian stocks while
the third one by American stocks.

Another interesting observation is that all three eigenvector components for
some of the Asian emerging markets such as China,
India, Pakistan, and Taiwan are too small
to be detected. That means that in the first approximation these indices are not influenced by
the world index dynamics at all. In Europe we saw no such correlation-free countries. However,
the eigenvector components of Greece, Portugal, Russia, and Turkey were somewhat smaller than
those of other European stock indices. North and South american stock indices have
approximately equal components with, perhaps, only Peru and
Venezuela somewhat falling behind.

It is interesting to compare our findings to those which were
previously obtained for {\it weeekly} returns using the
Minimal Spanning Tree (MST) technique \cite{bonanno}.
In this work it was also observed that indices are strongly grouped by the region
with most indices of, say, Asian stock indices forming a separate
branch of the MST. The market indices of Turkey, Greece, India, and
Pakistan were found to be weakly correlated with the
world index in both Ref. \cite{bonanno} and our study.
We believe that the spectral analysis method, used in our
work, gives a complimentary picture of reproducible correlations contained
in the matrix $C_{ij}$ to that of the MST method. One of the strong
points about the spectral analysis method is that it gives a clear
quantitative criterion for separating the signal from the
noise. Also with just a few large eigenvalues and their corresponding
eigenvectors the output of the spectral analysis is easier
to interpret and compare between, say, different time windows than that
of the MST technique.

The leading eigenvector component of a stock market index of a given country
can serve as a rough measure of the level of globalization of the financial sector of
this country. This point of
view is illustrated in Fig. 3 where the largest eigenvalue component is plotted
as a function of the rank of the country in the A.T. Kearney/Foreign Policy
Magazine Globalization Index$^{\rm TM}$ \cite{kearney}.
One can see a clear correlation
between high globalization rank (1 being the highest and 50 the smallest) and the
leading eigenvector component.

%\section{Influence of the time horizon on the correlation spectrum}

We further decided to explore how the outcome of the above eigenvector/eigenvalue
analysis dependent on the time horizon $\Delta t$ over which one computes the
returns of an index. When instead of daily open-to-close
returns we repeated our analysis for one day close-to-close returns the largest
eigenvalues have changed from $2.2, 3.5$, and $8.7$ to $2.1$, $3.7$, and $11.0$.
A noticeable 25\% increase in the largest eigenvalue perhaps acn be attributed to
markets having more time to respond to the news. Also, while daily
open-to-close returns in Asia, Europe, and Americas have almost no overlap (i.e.
time when two markets are simultaneously open), this situation is improved
when one considers daily close-to-close fluctuations. As shown in Fig. 4, the
largest eigenvalue continued to grow (albeit slowly), as the time
horizon of close-to-close returns was changed from one to ten trading days (weekly return
usually corresponds to just 5 trading days), reaching the value of $16.2$ for the longest
time horizon. However, the largest eigenvectors computed for these very different
time horizons remained remarkably stable. For example the average overlap between
the highest rank eigenvectors computed for these ten different time horizons
turned out to be $0.99$, i.e. these eigenvectors on average are only $1\%$
different from each other! Overlaps were somewhat smaller for lower ranking
eigenvectors with average values of $0.77$ for the second and $0.61$ for the
third largest eigenvalues. Still, as can be seen in the inset to Fig. 4, even in
the third eigenvector many of the main features are very robust with respect to
changes in the time horizon.

In an attempt to establish how relevant are magnitudes (as opposed to signs)
of price fluctuations to the observed correlation patterns we have repeated
the above analysis using $\delta X_i(t)={\rm sign} (S_i(t+\Delta t)-S_i(t))$.
The observed eigenvectors for different time horizons had $0.99$ average
overlap with those computed using $\delta X_i(t)=\ln S_i(t+\Delta t)-\ln S_i(t)$.
The largest eigenvalue again grew with the time horizon
from $7.4$ for signs of one day close-to-close returns to $10.4$ for signs of
5-day close-to-close returns. Lower rank eigenvectors of the correlation matrix
of signs also had substantial overlaps of $0.93$ and $0.86$ with normal ones.
This allows us to conclude that it is {\it signs} and not magnitude of returns
which are mostly relevant for the observed correlation patterns.

%\section{Correlations between the returns on different trading days.}

In what was described above we always computed (nearly) synchronous
correlations of different market returns on the same day
(or the same week for longer time horizons).
One has to take into account that due to the time-zone
difference daily open-to-close returns computed on the same trading day
(say, February 13) are not actually synchronous, with Asian stock markets
in the lead, followed by European and later with a small overlap by
American markets. However, the significance
of this time zone difference for say weekly returns is much less
pronounced. To check if the predictability of daily returns
is restricted to the same trading day or survives for several days
we have investigated the correlations in daily close-to-close returns
with all Asian indices shifted by $S$ days. The negative
values of the shift $S$ corresponds to correlations of
daily returns of Asian indices $|S|$ days {\it after} an observed
pattern of returns of the European and American stocks, while
positive $S$ corresponds to Asian indices preceding the rest of
the world by $S$ days. The simplest way to detect the
presence of correlations in this case is by calculating
the average correlation coefficient connecting any of the
$14$ Asian indices with $15+8=23$ European and American ones.
The size of the sample over which this average
is taken is $14 \times 23=322$. To check for reproducibility
of the observed patterns we divided our data set consisting of $226$ trading days
into three equal length segments and calculated this average for each segment independently.
The results are shown in the main panel of Fig. 5.
Reproducible correlations seem to survive for up to $3$ days on the
negative part of the axis, corresponding to the reaction of Asian
stocks to changes in European and American indices. At least
the sign of the average correlation coefficient was found to be
consistently positive in all three of our segments. $S=-1$ has the
largest magnitude of correlations. These (positive) correlations
represent the next trading day reaction of Asian stocks to changes
in European and American ones. Due to time-zone differences
a somewhat smaller (yet still positive) correlation coefficients
observed for $S=0$ correspond to the opposite effect, i.e the
response of European and American stocks to the events in Asia on
the day before. On the positive $S$ side there seems to be
a reproducible negative correlation at $S=1$ followed by a noisy
signal for larger values of $S$. Our results for $S=0$ and $S=1$
are in agreement with positive correlations between open-to-close
returns at Tokyo and New York stock exchanges that were previously reported
in the economics literature \cite{econo-us-japan} as well as in
recent econophysics papers \cite{vandewalle}. The inset to Fig. 5 show
the results of the same analysis for the whole $226$-day interval
when either Asian, or European and Asian stocks are shifted by $S$ days.
It is interesting that in this graph which has 3 times better statistics than
the main panel of the figure, there seem to be oscillations
of correlation coefficients with period of two trading days
for negative values of $S$ .

Work at Brookhaven National Laboratory was carried out under
Contract No. DE-AC02-98CH10886, Division of Material Science,
U.S.\ Department of Energy.

\onecolumn
\widetext

\begin{figure}
\centerline{
\epsfxsize=6in
\epsffile{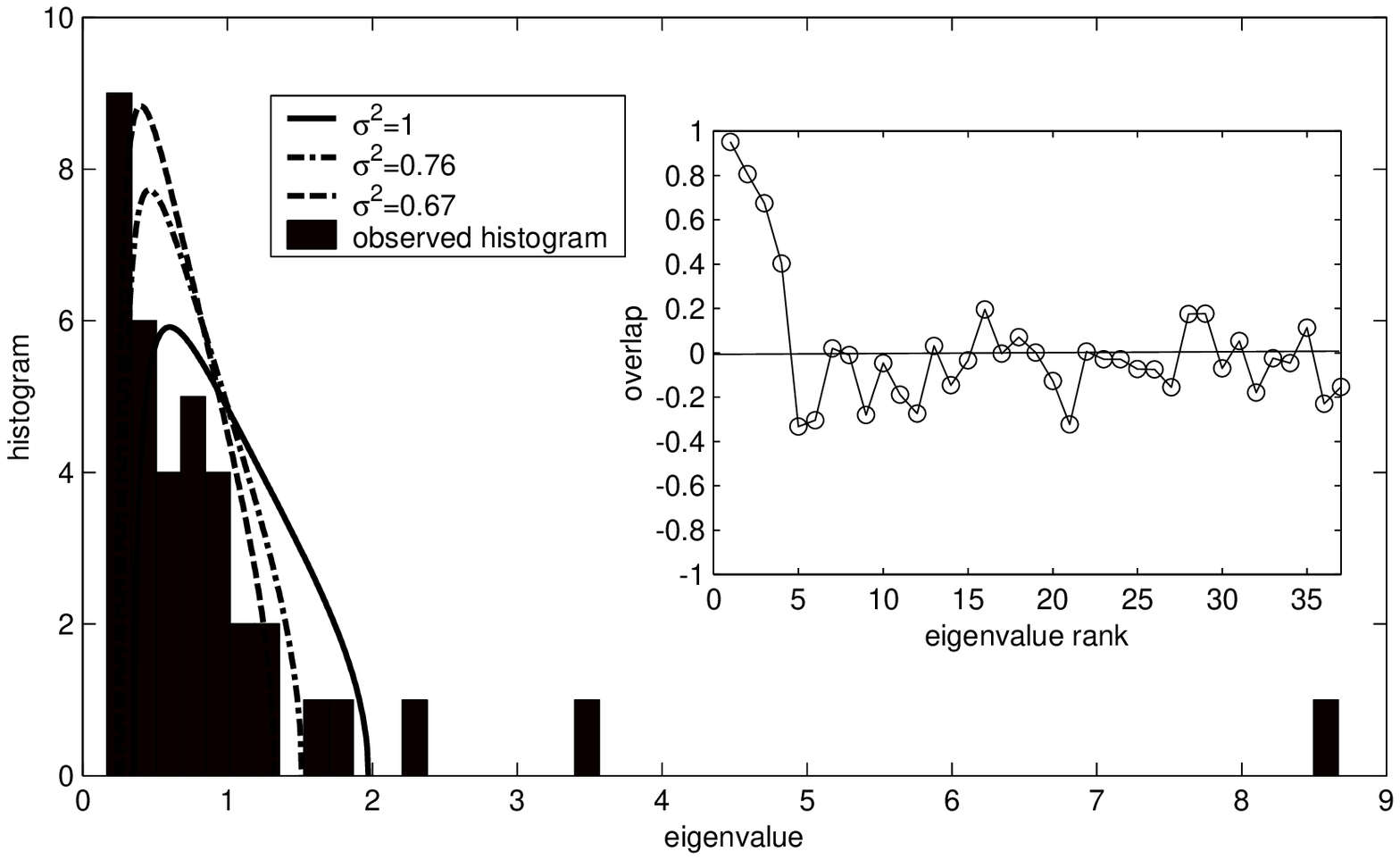}
}
\caption{The histogram of eigenvalues of the matrix $C_{ij}$. The straight
line is a fit with the Eq. (1) using $T=226$, $N=37$, and various values of $\sigma$,
corresponding to the exclusion of fluctuations contained in the largest
($\sigma^2=0.76$), or two largest ($\sigma^2=0.67$) eigenvalues. The inset shows
the overlap of eigenvectors computed for two consecutive $113$-day intervals
as a function of the rank of an eigenvalue. The three (perhaps even four) leading
eigenvectors clearly have a higher than random overlap.}
\end{figure}

\newpage
\begin{figure}
\centerline{\epsfxsize=6in
\epsffile{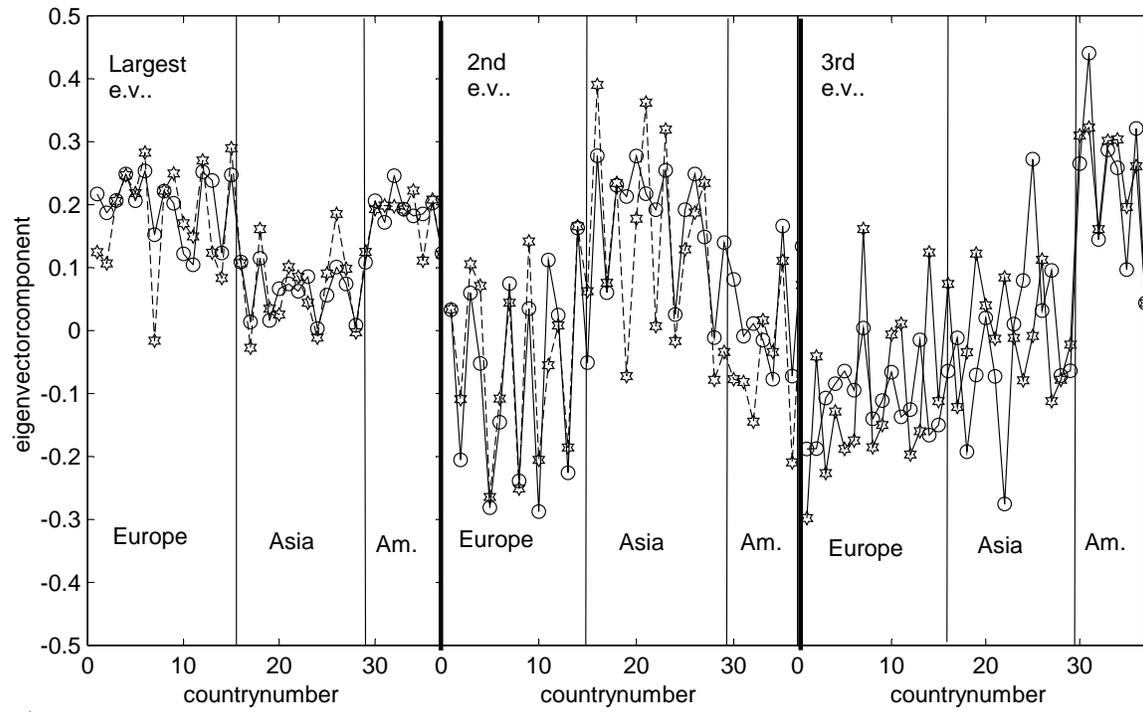}
}
\caption{(Components of the three leading eigenvectors
calculated for two consecutive $113$-day intervals
as a function of the country number (see Table 1.)}
\end{figure}
\newpage
\begin{figure}
\centerline{\epsfxsize=6in
\epsffile{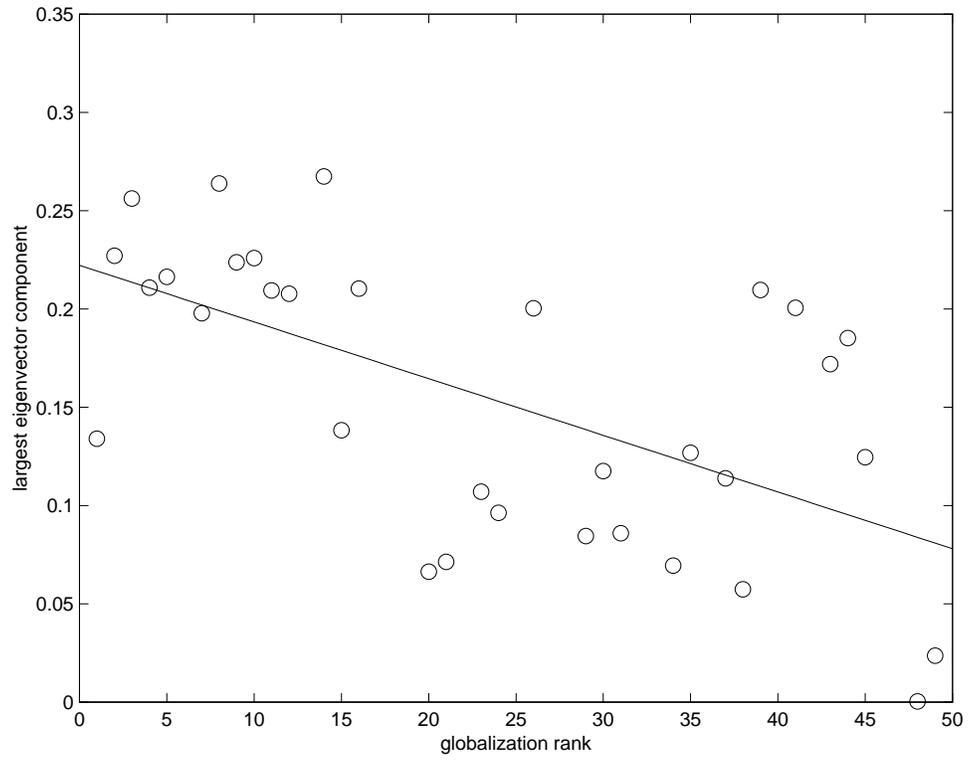}
}
\caption{The component of the highest ranking eigenvector
as a function of the globalization rank of the country from
Ref.[1]. The straight line is a linear fit to the data.}
\end{figure}

\begin{figure}
\centerline{\epsfxsize=6in
\epsffile{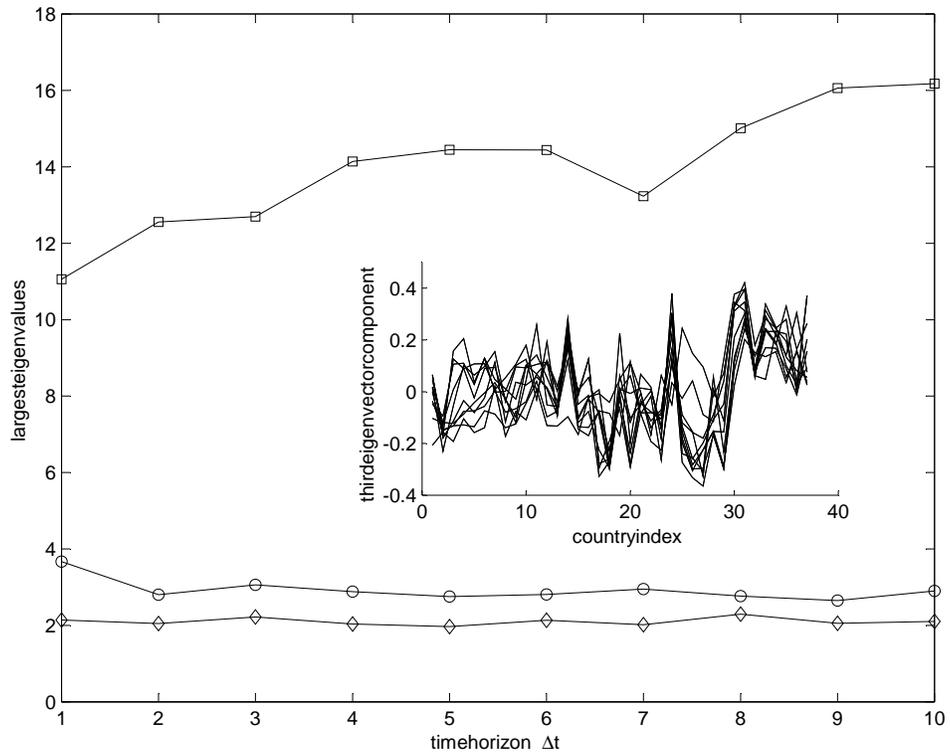}
}
\caption{Three largest eigenvalues of the correlation matrix
of close-to-close returns as a function of the time horizon (number on days
used to calculate the returns). The inset shows the components of the third
eigenvector for all ten time horizons.}
\end{figure}

\begin{figure}
\centerline{\epsfxsize=6in
\epsffile{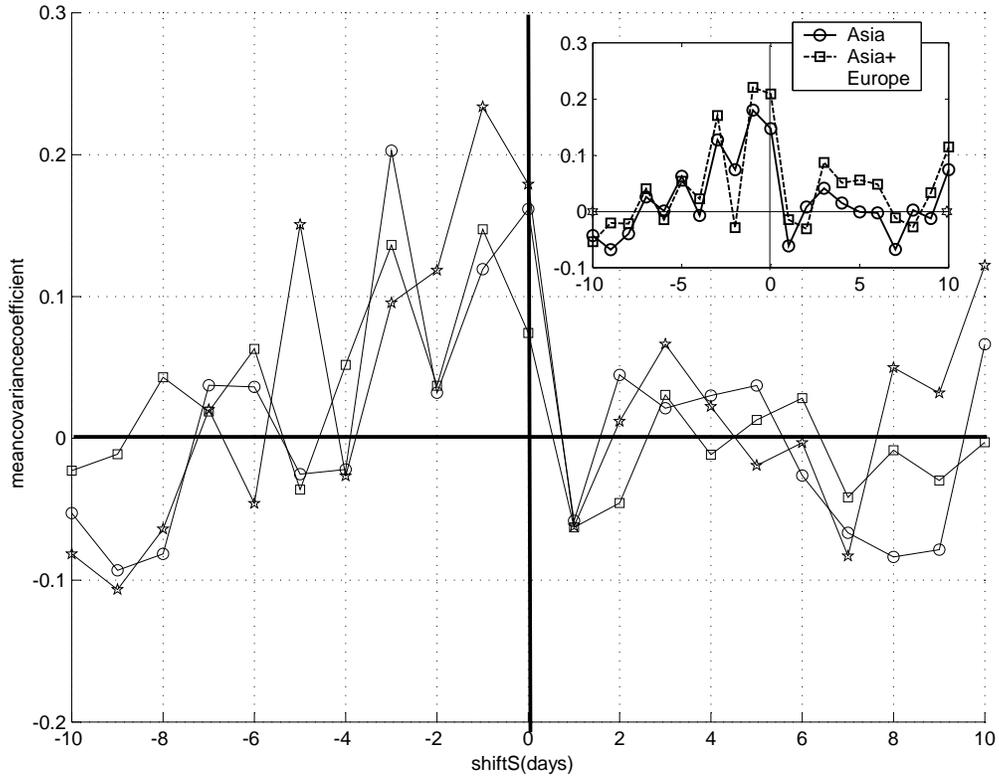}
}
\caption{The mean value of the correlation coefficient connecting Asian indices to the
rest of the world as a function of the shift $S$. Negative values of $S$ correspond to Asian
indices taken $|S|$ days later than the rest of the world. Three data sets were taking in three
equal length subintervals of our data set. The inset shows the same analysis repeated
including all data points with Asian indices (circles) and Asian and European indices (squares)
shifted. Note oscillations for negative $S$.}
\end{figure}

%\onecolumn
%\widetext

\begin{table}
\centerline{\epsfxsize=6in
\epsffile{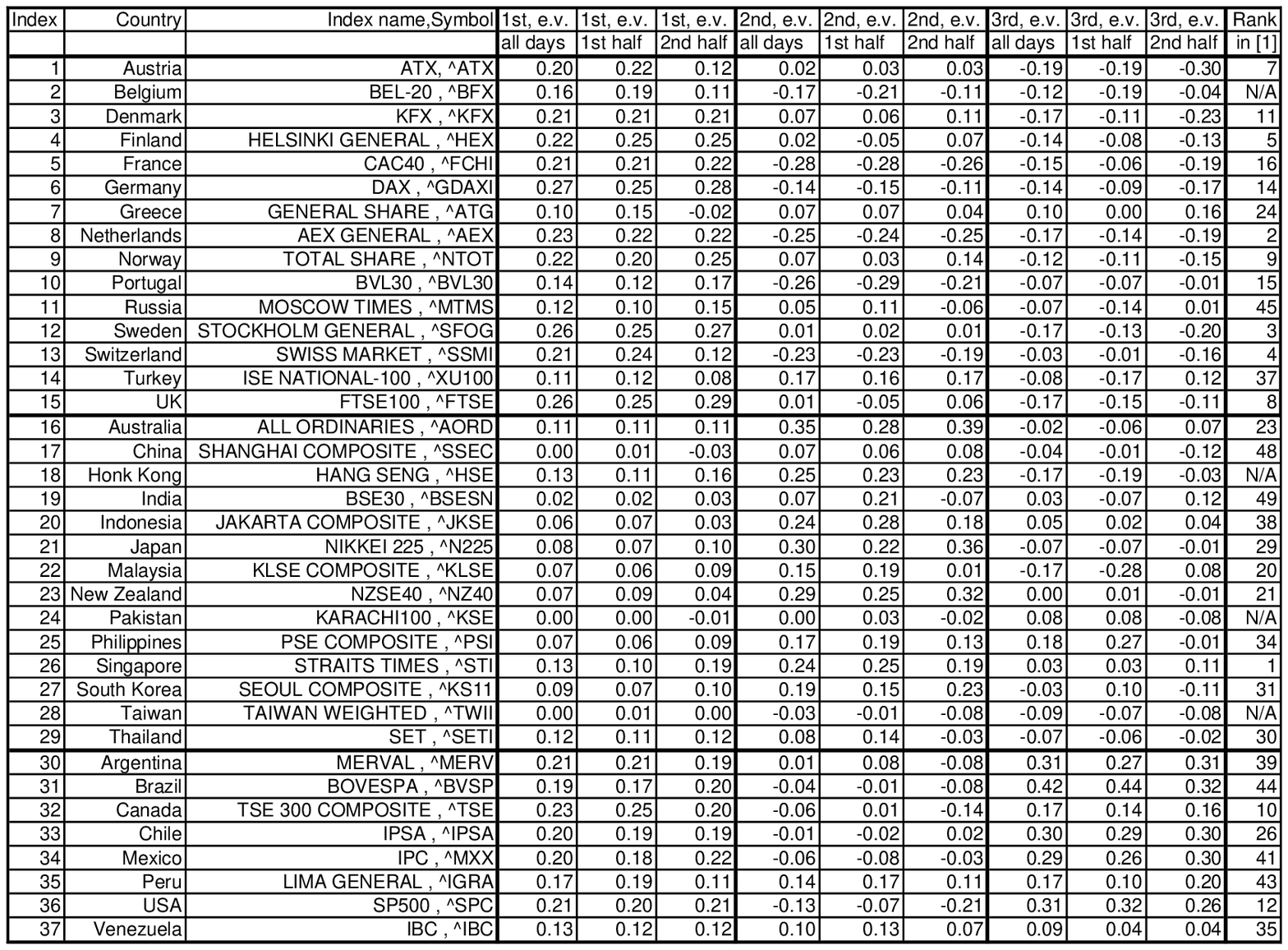}
}
\caption{Components of three leading eigenvectors computed for the whole
$226$-day time interval and its first and second halves. The last column is the
rank of the globalization index of the country as defined in Ref. [1] }
\end{table}

\end{document}